\documentclass[aps,twocolumn,pre,showpacs]{revtex4}
\usepackage{graphicx}
\usepackage{epsfig}
\usepackage{amsmath}
\usepackage{graphics}
\begin{document}

\title{Calculations of   canonical averages  from the  grand canonical ensemble
}

\author{D. S. Kosov$^1$}
\author{M. F. Gelin$^1$}
\author{A.I. Vdovin$^2$}

\affiliation{$^1$Department of Chemistry and Biochemistry, University of Maryland, College Park, MD  20742, USA}
\affiliation{$^2$Bogoliubov Laboratory of Theoretical Physics, Joint Institute for Nuclear Research, 141980 Dubna,
Moscow region, Russia}

\begin{abstract}
Grand canonical and canonical ensembles become equivalent in the thermodynamic limit,
but when the system size is finite the results obtained in the two ensembles deviate from each other.
In many important cases, the canonical ensemble provides
an appropriate physical description but it is often much easier to perform
the calculations in the corresponding grand canonical ensemble.
We present a method to compute  averages in canonical ensemble based
on calculations of the expectation values in grand canonical ensemble.
The number of particles, which is fixed in the canonical ensemble,
is not necessarily the same as the average number of particles  in the grand canonical ensemble.
\end{abstract}


\maketitle

\section{Introduction}
Whenever we need to work with a system of many quantum particles
it is much easier to perform the calculation in the grand canonical ensemble
than the corresponding calculations in the canonical ensemble.
For example, the calculations of the grand canonical partition function for ideal gas of
fermions or bosons is trivial, whereas the computation of the canonical partition
function for the the same system becomes a formidable  task even for a small number of particles.
Provided that the system is thermodynamically large, the canonical and
grand canonical descriptions agree with each other.
There are many quantum systems where the canonical description is more appropriate,
these include hot nuclei \cite{rossignoli94}, ultrasmall metallic grains\cite{delft},
Bose and Fermi gases in atomic traps \cite{politzer96,herzog97} and atoms in plasmas\cite{gilleron}.
Therefore, it is important to have a practical theoretical and computational method which enables us
to extract canonical averages from corresponding grand canonical calculations.

The problem, which we would like to solve in this paper is the
following. Suppose that we  can perform calculations  (or
measurements) in grand canonical ensemble which  is characterized
by temperature $T$ and chemical potential $\mu$. We would like to
compute expectation values of an arbitrary quantity $O$ in the
canonical ensemble with  temperature $T$ and the number of
particles $n$ by using only the averages in the grand canonical
ensemble. The number of particles $n$, which is fixed in canonical
ensemble, is not necessarily  the same as the average number of
particles  $\langle N \rangle$ in the grand canonical ensemble.

Earlier works on particle number projection in grand canonical
ensembles have been performed
to treat hot nuclei\cite{rossignoli94,rodriguez05,tanabe05,nakada06}, Bose-Einstein
condensation\cite{politzer96,herzog97,fujiwara70}
and to formulate canonical statistical mean field approximation for mesoscopic systems\cite{ ponomarenko}.
Our approach is
very different. First, we do not rely on projection operators but
extract  information from the grand canonical averages by
inverting fluctuation matrix. Second, $\langle N \rangle$ does not
necessarily equal to $n$ although it can be.

\section{Theory}
A quantum mechanical system has the Hamiltonian $H$. The Hamiltonian $H$
commutes with the number of particles operator $N$
\begin{equation}
[H,N]=0.
\end{equation}
Therefore the Hamiltonian and the number of particles operator have the same set of eigenvectors:
\begin{equation}
N |n\alpha \rangle =n |n \alpha \rangle,
\end{equation}
\begin{equation}
H |n\alpha \rangle =E_{n\alpha} |n\alpha \rangle.
\end{equation}
Let $O$ be an arbitrary operator.
The average in the grand canonical ensemble is
\begin{equation}
\langle O \rangle =\frac{1}{Z} \sum_{n \alpha}
e^{-\beta(E_{n\alpha} -\mu n)} \langle n\alpha |O|n \alpha
\rangle, \label{gca}
\end{equation}
with $Z$ being the grand canonical partition function
\begin{equation}
Z=\sum_{n \alpha} e^{-\beta(E_{n\alpha} -\mu n)},
\label{gcpart}
\end{equation}
and $\beta=1/kT$.
The average in canonical ensemble is
\begin{equation}
\langle O \rangle_n=\frac{1}{Z_n} \sum_{ \alpha} e^{-\beta
E_{n\alpha} } \langle n\alpha |O|n \alpha \rangle,
\end{equation}
with $Z_n$ being the  canonical partition function
\begin{equation}
Z_n=\sum_{ \alpha} e^{-\beta E_{n\alpha} }.
\end{equation}
We can consider $\langle O \rangle_n$ as a function of $n$ and
expand it as the power series:
\begin{equation}
\langle O\rangle_n=\sum_{k=0}^{\infty} q_k n^k.
\end{equation}
With this expansion the grand canonical average (\ref{gca}) becomes (appendix A):
\begin{equation}
\langle O \rangle = \langle \sum_{k=0}^{\infty} q_k N^k \rangle.
\end{equation}
Next, we take canonical expectation value and add/subtract the grand canonical  average
\begin{equation}
 \langle O\rangle_n= \sum_{k=0}^{\infty} q_k n^k + \langle O\rangle - \langle
\sum_{k=0}^{\infty} q_k N^k \rangle.
\end{equation}
Then we cancel $k=0$ terms in the both sums and the canonical expectation value becomes
\begin{equation}
\langle O \rangle_n= \langle O \rangle  -\sum_{k=1}^{\infty} q_k
(\langle N^k \rangle -n^k). \label{expansion}
\end{equation}
This  expression for canonical average of the operator $O$ involves
only grand canonical expectation values.
The coefficients $q_k$ are yet to be calculated. To determine them we introduce the operator
\begin{equation}
\bar{O}= O-  \sum_{k=1}^{\infty} q_k N^k
\label{baro}
\end{equation}
 and   compute the expectation value
$ \langle\bar{O} f(N) \rangle$, where $f(N)$ is  an arbitrary
function of $N$. We show in appendix A that this expectation value
can be split:
\begin{equation}
\langle  \bar{O}  f(N)\rangle = \langle \bar{O} \rangle \langle
f(N)\rangle. \label{eqforq}
\end{equation}
Since $f(N)$ is an arbitrary function of $N$, Eq.(\ref{eqforq}) is equivalent to the following system of equations
\begin{equation}
\left.
\begin{array}{c}
\langle   \bar{O}  N \rangle
= \langle \bar{O}\rangle \langle N\rangle \\
\langle    \bar{O}  N^2\rangle
= \langle \bar{O} \rangle\langle  N^2\rangle \\
.....
\\
\langle   \bar{O}  N^k\rangle
= \langle \bar{O} \rangle\langle  N^k\rangle \\
....
\end{array}
 \right\}
 \label{soe}
 \end{equation}
 If we use the explicit form for operator $\bar{O}$  (\ref{baro}),  the system of equations (\ref{soe}) becomes:
 \begin{equation}
 \sum_{k=1}^{\infty} q_k A_{km} =\langle ON^m\rangle -\langle O\rangle \langle N^m\rangle,
 \label{soe2}
 \end{equation}
 where the matrix $A_{km}$  is built from the fluctuations
 \begin{equation}
 A_{km} =\langle N^{k+m}\rangle  -\langle N^k\rangle\langle N^m\rangle.
 \end{equation}
 The system of linear algebraic equations (\ref{soe2}) along with the expansion (\ref{expansion})
 represent the main result of the paper. The similar equations (although for the case
 $\langle N\rangle=n$) were obtained by the methods of thermofield dynamics in the context of
 superconducting nuclei at finite temperature\cite{kosov96a,kosov96b}.

It can be readily shown by the direct differentiation of the partition function
(\ref{gcpart}) that
\begin{equation}
 A_{km} =\frac{1}{Z^2 \beta^{k+m}} \left( Z \frac{\partial^{k+m} Z}{\partial \mu^{k+m}} -
  \frac{\partial^{k} Z}{\partial \mu^{k}}  \frac{\partial^{m} Z}{\partial \mu^{m}}
 \right).
 \label{adir}
 \end{equation}

Now we prove the convergence of the expansion (\ref{expansion}).
It is possible to demonstrate based on simple consideration  that 
 the difference between  canonical and grand canonical averages is \cite{mcquarrie}
\begin{equation}
\langle O \rangle - \langle O \rangle_n \sim \frac{1}{\langle N \rangle}.
\label{o-on}
\end{equation}
The calculation in appendix C  shows that for $k>1$
\begin{equation}
 q_{k}\sim1/\left\langle N\right\rangle ^{k}.
 \label{qscale}
 \end{equation}
 Suppose that $n=\langle N \rangle$. Then  $(\langle N^k \rangle -n^k)
  \sim \langle N \rangle^{k-1}$ for large $\langle N \rangle$.
  Therefore, the corrections to the grand canonical average in Eq.(\ref{expansion}) become in this case
\begin{equation}
\sum_{k=1}^{\infty} q_k
(\langle N^k \rangle -n^k) \simeq \frac{1}{\langle N \rangle} \sum_{k=1}^{\infty} c_k ,
\label{limit}
\end{equation}
where $c_k$ are some coefficients.
Comparing (\ref{o-on}) and (\ref{limit}) we see that $\sum_{k=1}^{\infty} c_k$ is finite. It means that 
the expansion (\ref{expansion}) converges when $\langle N \rangle =n$.
We would like to note at this point that each term in the sum (\ref{expansion}) is 
$\sim 1/\langle N \rangle$.  It means that the method becomes computationally efficient when it is applied to the large systems, since one needs to include less terms in the expansion to achieve the same accuracy  in this case.
Let us consider the corrections to the grand canonical average in Eq.(\ref{expansion}) 
for the case $n=\langle N \rangle +j$ where $j$ is some integral number 
\begin{eqnarray}
\langle O \rangle - \langle O \rangle_n =
\sum_{k=1}^{\infty} q_k
(\langle N^k \rangle -[\langle N \rangle +j]^k).
\label{j1}
\end{eqnarray}
We assume that $ j/\langle N \rangle \ll1 $. If we substitute the expression for 
$\langle N^k \rangle$ (\ref{c17}) into (\ref{j1}) and use the binomial expansion
up to the first order in $j/\langle N\rangle$  for $[\langle N \rangle +j]^k$, we obtain 
\begin{equation}
\langle O \rangle - \langle O \rangle_n =
\sum_{k=1}^{\infty} q_k \langle N \rangle^{k-1} \frac{c k (k-1)}{2} \left [1- \frac{2j}{c (k-1)} \right].
\label{series}
\end{equation}
The series (\ref{series}) always converges for $j=0$ as we just demonstrated. To prove the convergence for $j\ne0$ we split the sum (\ref{series}) into two parts: the first part is  for $0 < k \le K$ and the second part is for $K<k< \infty$, where $K$ is some positive integer.  The first part  is always finite and $\sim 1/ \langle N \rangle $. By choosing $K$, we can always make 
$\left [1- \frac{2j}{c (k-1)} \right] \simeq 1$ for $k>K$. Therefore, the convergence of the 
expansion (\ref{expansion}) for $\langle N \rangle =n$ case implies the convergence for any finite 
difference $|\langle N \rangle -n |$ for which $\frac{|\langle N \rangle -n |}{\langle N \rangle} \ll 1$.
 We note that these arguments may not work when the matrix $A_{nm}$ is  singular.
For example, in the case of low temperature Fermi gas, all matrix
elements $A_{nm}$ tend to zero, therefore canonical and grand
canonical descriptions may deviate from each other in the
thermodynamic limit due to the persistent existence of a
few-particle fluctuations in the grand canonical ensemble
\cite{bowen}.

\section{Example calculations}

As an example we consider the system of noninteracting quantum particles distributed on
$n_{levels}$ single particle energy levels with
energies $\varepsilon_{l}$. The logarithm of the grand canonical partition function is
\begin{equation}
\ln Z= \eta \sum_{l=1}^{n_{levels}} \ln\left( 1+ \eta e^{\beta(\mu-\varepsilon_l)} \right),
\label{fdpartition}
\end{equation}
where $\eta=+1$ is for fermions and $\eta=-1$ is for bosons.
 We set $\beta=1$, $\varepsilon_l=l$, and $n_{levels}=5$ in all our calculations.

\begin{table}
\caption{Occupation numbers and total energy for fermions. FD
refers to the Fermi-Dirac statistics in grand canonical ensemble
($\langle N \rangle=4$). The number of particles in the canonical ensemble is 2. $k_{max} $ is the number of terms
included in expansion (\ref{expansion}). }
\begin{tabular}{ccccccccc}
\hline
\hline
 &                           &       & \multicolumn{5}{c}{$k_{max}$} & \\
 l & $\varepsilon_l$ & FD &  1 & 2 & 3 & 4 & 6 & exact\\
\hline
1 & 1.0 &  0.98  &  0.92   & 0.87  & 0.86 & 0.87 & 0.87 & 0.87 \\
2 & 2.0 &  0.95  & 0.80    & 0.69  & 0.67 & 0.67 & 0.68 & 0.68 \\
3 & 3.0 &  0.87  & 0.52    & 0.34  & 0.34 & 0.30 & 0.30 & 0.30 \\
4 & 4.0 &  0.72  & 0.07    & 0.01  & 0.06 & 0.11 & 0.12  & 0.12 \\
5 & 5.0 &  0.48  & -0.31   & 0.10  & 0.06 & 0.04 & 0.04 & 0.04 \\
\hline
\multicolumn{2}{c}{total energy} & 10.77   & 2.82 &  3.77&  3.78   & 3.79   & 3.79 & 3.79  \\
\hline
\end{tabular}
\end{table}

\begin{table}
\caption{Occupation numbers and total energy for bosons. BE refers
to the Bose-Einstein statistics in grand canonical ensemble
($\langle N \rangle=4$). The number of particles in the canonical ensemble is 2. $k_{max} $ is the number of terms
included in expansion (\ref{expansion}). }
\begin{tabular}{ccccccccc}
\hline
\hline
 &                           &       & \multicolumn{5}{c}{$k_{max}$} & \\
 l & $\varepsilon_l$ & BE &  1 & 3 & 5 & 7 & 11 &  exact\\
\hline
1 & 1.0 & 3.43 & 1.52  & 1.52  & 1.47 & 1.44 & 1.42  & 1.42 \\
2 & 2.0 & 0.40 & 0.33  & 0.33  & 0.37 & 0.39  & 0.39 & 0.39 \\
3 & 3.0 & 0.12 & 0.10  & 0.10  & 0.11  & 0.12  & 0.13 & 0.13 \\
4 & 4.0 & 0.04 & 0.04  & 0.04  & 0.04  & 0.04  & 0.05 &  0.05 \\
5 & 5.0 & 0.01 & 0.01  & 0.01  & 0.01  & 0.02  & 0.02&   0.02 \\
\hline
\multicolumn{2}{c}{total energy} & 4.81& 2.68  & 2.69 &  2.77&  2.82  & 2.84 &  2.85  \\
\hline
\end{tabular}
\end{table}

\begin{table}
\caption{Occupation numbers and total energy for fermions. FD
refers to the Fermi-Diract statistics in grand canonical ensemble
($\langle N \rangle=2$).   The number of particles in the
canonical ensemble is 4. $k_{max} $ is the number of terms
included in expansion (\ref{expansion}). }
\begin{tabular}{ccccccccc}
\hline
\hline
 &                           &       & \multicolumn{5}{c}{$k_{max}$} & \\
 l & $\varepsilon_l$ & FD &  1 & 2 & 3 & 4 & 5 & exact\\
\hline
1 & 1.0 &  0.80  & 1.18   & 0.92  & 0.98 & 0.99 & 0.99 & 0.99 \\
2 & 2.0 &  0.60  & 1.18   & 1.01  & 0.97 & 0.93 & 0.97 & 0.97 \\
3 & 3.0 &  0.36  & 0.91   & 1.02  & 0.96 & 0.97 & 0.91 & 0.91 \\
4 & 4.0 &  0.17  & 0.51   & 0.70  & 0.72 & 0.73 & 0.77  & 0.77 \\
5 & 5.0 &  0.07  & 0.23   & 0.35  & 0.38 & 0.37 & 0.36  & 0.36 \\
\hline
\multicolumn{2}{c}{total energy} & 4.10   & 9.42 &  10.53&  10.54   & 10.55   & 10.55 & 10.55  \\
\hline
\end{tabular}
\end{table}

First, we extract averages in canonical ensemble for the smaller
system from  grand canonical ensemble for the larger system. We
select the chemical potential $\mu$ in such a way that the average
number of particles $\langle N\rangle$
 in the grand canonical ensemble is 4.
We would like to extract the information about the canonical
ensemble with $n=2$ particles from this grand canonical ensemble.
We compute the occupation numbers and then all physical quantities
like total energy can be calculated with the use of these
occupation numbers. To start our calculations we set $O=n_{l}$,
where $n_l$ is the operator of the  number of particles on level
$l$. Then we solve the linear system of linear equations
(\ref{soe2}) to find the coefficients $q_k$ and use these $q_k$s
to calculate the grand canonical occupation numbers by
Eq.(\ref{expansion}). The matrix elements $A_{nm}$  are computed
by Eq.(\ref{adir}) and with  the help of recurrent relation from
appendix B. The matrix element in the right hand side of
Eq.(\ref{soe2}) is computed as the following derivative of the
grand canonical partition function (\ref{fdpartition}):
\begin{equation}
 \langle n_l N^m \rangle =- \frac{1}{Z \beta^{m+1}} \frac{\partial^{m+1} Z}{\partial \varepsilon_l \partial \mu^{m}} .
 \label{dirme}
  \end{equation}
The results of these calculations are shown in Table I (fermions)
and Table II (bosons). The fermionic and bosonic systems both show
the convergence to the exact results as we include more terms in
expansion (\ref{expansion}). The convergence for bosons is slower
than that for fermions. It is due to the fact that the
fluctuations of the occupation numbers $\langle \Delta n_l^2
\rangle = \langle n_l \rangle - \eta \langle n_l \rangle^2$ tend
to be larger for bosons ($\eta=-1$) than for fermions ($\eta=+1$).

The method also works in the opposite direction,  therefore we can
compute averages in canonical ensemble for the larger system using
grand canonical averages for  the smaller system. We select the
grand canonical ensemble with  $\langle N\rangle=2$ and  we
compute the occupation numbers in the canonical ensemble of $n=4$
particles. Table III shows the results of these calculations for
noninteracting fermions.  The convergence to the exact values is
as good as in the previous case, thereby it demonstrates that the
method can be also used to extract the canonical ensemble
information for the larger system from the grand canonical
calculations of the smaller system. The very similar results were
obtained for bosons and we do not show it here.

\section{Conclusions}
We formulated the method to compute  averages in canonical ensemble based on calculations
in grand canonical ensemble. The number of particles $n$, which is fixed in the canonical ensemble,
is not necessarily the same as the average number of particles  $\langle N \rangle$ in the grand canonical ensemble.
Expansion (\ref{expansion})  and system of linear algebraic equations (\ref{soe2}) for coefficients
of the expansion  are  the main result of the paper.
We performed the test calculations for ideal Fermi and Bose gases,
compared our calculations with the exact results and demonstrated convergence
properties of expansion (\ref{expansion}).

\begin{acknowledgments}
 This work has been supported by NSF-MRSEC
DMR0520471 at the University of Maryland and   by the American
Chemical Society Petroleum Research Fund (44481-G6).
\end{acknowledgments}

\appendix
\section{Useful identities}
\begin{widetext}
Proof that if $\langle O\rangle_n = \sum_{k=0}^{\infty} q_k n^k $
then $\langle O \rangle= \langle \sum_{k=0}^{\infty} q_k N^k
\rangle$.
\begin{eqnarray}
\langle  O\rangle  =\frac{1}{Z} \sum_{n \alpha}
e^{-\beta(E_{n\alpha} -\mu n)}\langle n\alpha |O|n\alpha\rangle
=\frac{1}{Z} \sum_{n }  e^{\beta \mu n} \sum_{\alpha}  e^{-\beta
E_{n\alpha} } \langle n\alpha |O |n\alpha \rangle =\frac{1}{Z}
\sum_{n } e^{\beta \mu n} Z_n \langle O\rangle_n \nonumber
\\
=\frac{1}{Z}  \sum_{n }  e^{\beta \mu n}  Z_n \sum_{k=0}^{\infty}
q_k n^k =\frac{1}{Z}  \sum_{n \alpha}  e^{-\beta(E_{n\alpha} -\mu
n)}    \sum_{k=0}^{\infty} q_k n^k =\frac{1}{Z}  \sum_{n \alpha}
e^{-\beta(E_{n\alpha} -\mu n)}   \langle n \alpha|
\sum_{k=0}^{\infty} q_k N^k | n \alpha \rangle =\langle
\sum_{k=0}^{\infty} q_k N^k \rangle \nonumber
\end{eqnarray}

Proof that $\langle   \bar{O}  f(N)\rangle = \langle \bar{O}
\rangle\langle f(N)\rangle $.

\begin{eqnarray}
\langle  \bar{O}  f(N)\rangle  =\frac{1}{Z} \sum_{n \alpha}
e^{-\beta(E_{n\alpha} -\mu n)}\langle n\alpha |\bar{O} f(N)|n
\alpha\rangle = \frac{1}{Z}  \sum_{n \alpha} e^{-\beta(E_{n\alpha}
-\mu n)} \langle n\alpha |\bar{O} |n\alpha\rangle f(n) \nonumber
\\
=\frac{1}{Z}  \sum_{n }  e^{\beta \mu n} f(n) \sum_{\alpha}
e^{-\beta E_{n\alpha} } \langle n\alpha |\bar{O} |n\alpha \rangle
=\frac{1}{Z} \sum_{n }  e^{\beta \mu n} f(n) \sum_{\alpha}
e^{-\beta E_{n\alpha} } \langle n\alpha |O-\sum_{k=1}^{\infty} q_k
N^k |n\alpha \rangle \nonumber
\\
=\frac{1}{Z}  \sum_{n }  e^{\beta \mu n} f(n) Z_n \left(\langle
O\rangle_n-\sum_{k=1}^{\infty} q_k n^k\right) =\frac{1}{Z} \sum_{n
}  e^{\beta \mu n} f(n) Z_n  q_0 = \langle f(N)\rangle  q_0 =
\langle f(N)\rangle \langle \bar{O}\rangle \nonumber
\end{eqnarray}
\end{widetext}

\section{Recurrent relation for the calculations of the derivatives}

We define $\partial^{m} = \partial^{m}/(\partial \mu )^{m}$.

Let
\begin{equation}
\Psi = \ln Z,\,\, Z_{m} = \partial^{m}Z.
\end{equation}
Then
\begin{equation}
Z_{m} = \partial^{m}(e^{\Psi}).
\end{equation}
Therefore
\begin{eqnarray}
Z_{m+1}=\partial^{m}\partial(e^{\Psi})=\partial^{m}(\Psi'e^{\Psi})
\nonumber
\\
=\sum_{k=0}^{m}\frac{m!}{k!(m-k)!}Z_{k}\Psi^{(m-k+1)},\label{1}
\label{b3}
\end{eqnarray}
where
\begin{equation} \Psi^{(k)}  =
\partial^{k}\Psi.
\end{equation}
Explicitly,
\begin{equation}
\Psi^{(0)} = \Psi = \eta \sum_{l=1}^{n_{levels}}\ln B_{l},\,\,\,
B_{l} = (1+\eta\exp\{\beta(\mu-\varepsilon_{l})\}).\label{Psi0}
\end{equation}
$\eta=1$ for fermions and $=-1$ for bosons.

Assume that
\begin{equation}
\Psi^{(k)}=\sum_{l=1}^{n_{levels}}\,\,\sum_{\sigma=0}^{k}a_{\sigma}^{k}(B_{l})^{-\sigma}.\label{2}
\end{equation}
Since
\begin{equation}
\partial(B_{l})^{-\sigma}=-\sigma\beta\{(B_{l})^{-\sigma}-(B_{l})^{-\sigma-1}\},
\end{equation}
we get then
\begin{equation}
\Psi^{(k+1)} = \partial\Psi^{(k)}=
\sum_{l=1}^{n_{levels}}\,\,\sum_{\sigma=0}^{k}-\sigma\beta
a_{\sigma}^{k}\{(B_{l})^{-\sigma}-(B_{l})^{-\sigma-1}\}.
\end{equation}
Since, according to (\ref{2})
\begin{equation}
\Psi^{(k+1)}=\sum_{l=1}^{n_{levels}}\,\,\sum_{\sigma=0}^{k+1}a_{\sigma}^{k+1}(B_{l})^{-\sigma},
\end{equation}
we find that
\begin{equation}
a_{\sigma}^{k+1}=-\beta(\sigma a_{\sigma}^{k}-(\sigma-1)a_{\sigma-1}^{k}).
\end{equation}
Evidently,
\begin{equation}
a_{0}^{1}=\beta,\,\, a_{1}^{1}=-\beta.
\end{equation}

\section{Scaling with the number of particles}

We use notations (B1)-(B5) from appendix B. In these notations
\begin{equation}
\left\langle N^{m}\right\rangle \equiv\frac{Z_{m}}{Z\beta^{m}}.
\end{equation}
Plugging this into Eq.(\ref{b3}) we get:
\begin{equation}
\left\langle N^{m+1}\right\rangle =
\sum_{k=0}^{m}\frac{m!}{k!(m-k)!}\left\langle N^{k}\right\rangle
\frac{\Psi^{(m-k+1)}}{\mathbf{\beta}^{m-k+1}}. \label{c3}
\end{equation}
In the thermodynamic limit, the sum over $l$ in (\ref{Psi0}) can
be replaced by the integral. Since the corresponding density of states
is proportional to the system volume $V$, e.g. in three dimensional space it becomes ($\mu$ is the mass of the particle)
\begin{equation}
\rho(\varepsilon)=\frac{V}{\sqrt{2}\pi^{2}}\left(\frac{\sqrt{\mu}}{\hbar}\right)^{3}\sqrt{\varepsilon},
\end{equation}
we see that $\Psi^{(m)}\sim V$. Since $\left\langle
N^{m}\right\rangle $ must be proportional to $V^{m}$, the term
with $k=m$ in Eq.(\ref{c3}) gives the leading contribution.
Retaining in (\ref{c3}) the two leading terms, we get:
\begin{equation}
\left\langle N^{m+1}\right\rangle \approx\left\langle N^{m}\right\rangle \left\langle N\right\rangle +
m\frac{\Psi^{(2)}}{\mathbf{\beta}^{2}}\left\langle N^{m-1}\right\rangle \label{ser1}
\end{equation}
with
\begin{equation}
\frac{\Psi^{(2)}}{\mathbf{\beta}^{2}}\equiv\left\langle
N^{2}\right\rangle - \left\langle N\right\rangle
^{2}\equiv\sum_{l}\left(\left\langle n_{l}\right\rangle
-\eta\left\langle n_{l}\right\rangle ^{2}\right).
\end{equation}
Here $\left\langle n_{l}\right\rangle $ are the BE or FD
occupation numbers.  Since $\left\langle N\right\rangle \gg1$, we
assume in Eq.(\ref{ser1}) that
\begin{equation}
\left\langle N^{m}\right\rangle =\left\langle N\right\rangle ^{m}+
\frac{\Psi^{(2)}}{\mathbf{\beta}^{2}}\alpha_{m}\left\langle N\right\rangle ^{m-2}+O(\left\langle N\right\rangle ^{m-4}).
\label{c5}
\end{equation}
Inserting (\ref{c5}) into (\ref{ser1}) and retaining the leading
terms, we get
\begin{equation}
\alpha_{m+1}=\alpha_{m}+m,\,\,\, \alpha_{m}=m(m-1)/2.\label{6}
\end{equation}
Thus
\begin{equation}
\left\langle N^{m}\right\rangle \approx\left\langle N\right\rangle ^{m}+
\frac{\Psi^{(2)}}{\mathbf{\beta}^{2}}\,\frac{m(m-1)}{2}\left\langle N\right\rangle ^{m-2}.
\label{nm}
\end{equation}
Now, we consider the terms $\left\langle n_{l}N^{m}\right\rangle
$. Following Eq.(\ref{dirme}) we differentiate Eq.(\ref{1}) over
$\varepsilon_{l}$, and  retaining only the leading terms in $V$,
we get:
\begin{equation}
\left\langle n_{l}N^{m+1}\right\rangle =\left\langle
n_{l}N^{m}\right\rangle  \left\langle N\right\rangle +\left\langle
N^{m}\right\rangle (\left\langle n_{l}\right\rangle
-\eta\left\langle n_{l}\right\rangle ^{2}).\label{nlm}
\end{equation}
We shall look for the solution in the form
\begin{eqnarray}
\left\langle n_{l}N^{m+1}\right\rangle =\left\langle n_{l}\right\rangle \left\langle N\right\rangle ^{m+1}
\nonumber
\\
+\gamma_{m+1}\left\langle N\right\rangle ^{m}+O(\left\langle N\right\rangle ^{m-1}).\label{5}
\end{eqnarray}
Inserting this equation into (\ref{nlm}), we obtain
$\gamma_{m+1}=\gamma_{m}+\left\langle n_{l}\right\rangle -\eta\left\langle n_{l}\right\rangle ^{2}$
and
\begin{eqnarray}
\left\langle n_{l}N^{m+1}\right\rangle \approx\left\langle n_{l}\right\rangle \left\langle N\right\rangle ^{m+1}
\nonumber
\\
+(m+1)(\left\langle n_{l}\right\rangle - \eta\left\langle
n_{l}\right\rangle ^{2})\left\langle N\right\rangle
^{m}.\label{nlm1}
\end{eqnarray}
Next, we consider the system of linear equations (\ref{soe2}). In
the thermodynamic limit, we retain only the first terms in
Eqs.(\ref{nlm1}) and (\ref{nm}).  If we apply a ``weak''
thermodynamic limit and retain the next-order terms in
Eqs.(\ref{nlm1}) and (\ref{nm}), then
\begin{equation}
A_{km}\approx km\frac{\Psi^{(2)}}{\mathbf{\beta}^{2}}\left\langle N\right\rangle ^{k+m-2},
\end{equation}
\begin{equation}
\left\langle n_{l}N\right\rangle ^{m}- \left\langle
n_{l}\right\rangle \left\langle N\right\rangle ^{m}\approx
m(\left\langle n_{l}\right\rangle -\eta\left\langle
n_{l}\right\rangle ^{2})\left\langle N\right\rangle
^{m-1}\label{8}
\end{equation}
Therefore, assuming that $\Psi^{(m)}\sim V \sim \langle N \rangle$ we get
\begin{equation}
q_{k}A_{km}\sim \left\langle
N\right\rangle ^{m-1},
\end{equation}
\begin{equation}
q_{k}\left\langle N\right\rangle ^{k+m-1}\sim\left\langle
N\right\rangle ^{m-1},
\end{equation}
therefore
\begin{equation}
q_{k} \sim 1/ \left\langle N \right \rangle^{k}.
\label{9}
\end{equation}
The same $\langle N \rangle$ dependence can be also obtained if we
express  the particle number operator $N$ in terms of
creation/annihilation operators and  apply Wick's
theorem\cite{wick} to matrix elements $\langle N^{m} \rangle$ and
$\langle n_l N^m\rangle$. 

Using the fact that $\Psi^{(2)} \sim \langle N \rangle $ we transform Eq.(\ref{nm}) to the form
\begin{equation}
\left\langle N^{m}\right\rangle \approx\left\langle N\right\rangle ^{m}+
 c \frac{m(m-1)}{2} 
\left\langle N \right\rangle ^{m-1}.
\label{c17}
\end{equation}
Here $c$ is some constant which does not depend on $m$ and $\langle N \rangle$.

It has not escaped our notice that Eqs.(\ref{ser1}) and
(\ref{nlm}) break down for bosons at critical and lower
temperatures, since, due to the Bose condensation, $\Psi^{(m)}$
becomes proportional to $V^{m}$.

\end{document}